\documentclass{article}

\usepackage{amsmath}
\usepackage{amsfonts}
\usepackage{amsthm}

\newcommand{\lb}{\left (}
\newcommand{\rb}{\right )}
\newcommand{\eq}[1]{\begin{equation} #1 \end{equation}}
\newcommand{\eqna}[1]{\begin{eqnarray} #1 \end{eqnarray}}

\newcommand{\p}{\partial}
\newcommand{\mF}{\mathcal{F}}
\newcommand{\e}{\epsilon}
\newcommand{\wk}{\widetilde{k}}
\newcommand{\ok}{\overline{k}}
\newcommand{\D}{\mathcal{D}}
\newcommand{\cmO}{\hat{\mathcal{O}}}
\newcommand{\wLambda}{\widetilde{\Lambda}}

\title {On relation between Nekrasov functions and BS periods in pure $SU(N)$ case}
\author {Popolitov A.\footnote{ {\small {\it
ITEP, Moscow, Russia}};
popolit@itep.ru}}

\begin {document}
	\maketitle
	
	\vspace{-6.0cm}

\begin{center}
	\hfill ITEP/TH-98/09\\
\end{center}

\vspace{4cm}

	\begin {abstract}
		We investigate the duality between the Nekrasov function and the quantized
		Seiberg-Witten prepotential, first guessed in \cite{GKMMM} and further elaborated
		in \cite {NS} and \cite{MMBS1}. We concentrate on providing more thorough checks
		than the ones presented in \cite {MMBS1} and do not discuss the motivation and historical
		context of this duality. 
		The check of the conjecture up to $o (\hbar^6, \ln (\Lambda))$ is done by hands
		for arbitrary $N$	(explicit formulas are presented). Moreover, details of 
		the calculation that are essential for the computerization of the check are worked out.
		This allows us to test the conjecture up to $\hbar^6$ and
		up to higher powers of $\Lambda$ for $N = 2,3,4$.
		Only the case of pure $SU(N)$ gauge theory is considered.
	\end {abstract}

	\section {Introduction}
	\par It is often realized during late decades that the quantities, calculated in two
	theories with different underlying mathematical machinery and/or physical origin,
	coincide. When this takes place one says that there is a duality between these
	two theories, and it often means that there is some sort of underlying structure or
	a unifying concept, which makes the duality evident.
	\par Discovering and investigating dualities is very important, since
	once an underlying concept is discovered it provides the right point of view for the both theories.
	Moreover, since some theories in modern mathematical physics still lack
	experimental evidence, the fact that they are dual to some other theories may serve
	as an indication that they do describe the real world. Moreover, even if the underlying
	concept is missing, the duality can be used to solve longstanding problems in one
	theory via using techniques from the other theory.

	\par Recently, the duality that connects a huge number of different theories in modern
	mathematical physics had been conjectured. In connects Seiberg-Witten theory \cite {GKMMM},
	\cite {SW1}-\cite {SWl}, the Nekrasov functions
	from the quiver theories \cite {Nf1}-\cite {Nfl}, conformal field theory \cite {CFT1}-\cite{CFTl}
	(the relation to CFT is provided by the celebrated AGT conjecture \cite {AGT1}-\cite {AGTl}),
	and matrix models in the Dijkgraaf-Vafa phase \cite {DV1}-\cite {DVl}. Importance of this unification,
	its traces being present in a number of new \cite{NS},\cite {MMBS1},\cite {AGT1}-\cite {AGTl},\cite{MMS}
	and less recent \cite {GKMMM},\cite {Wirc}-\cite {KS} papers, cannot be overestimated.
	However, although at conjectural level the whole picture is relatively clear \cite {MMS},
	checks and proofs still need to be done to be certain that this unification really takes place.
	
	\bigskip
	
	\par In this paper we are concentrating on a small part of the unification, that is, on
	the statement that the Nekrasov function with $\e_2 = 0$ is equal to the quantized SW prepotential
	provided $\e_2 = \hbar$. Furthermore, we restrict ourselves to the case of pure gauge theory,
	without matter hypermultiplets. The idea of this relation first appeared in \cite {GKMMM}
	and then was investigated in \cite {NS}, \cite {MMBS1}. However, for
	$SU(N)$ with $N > 2$ only simple checks were made (up to $o(\hbar^2, \Lambda^{2N})$).
	In this paper we present more thorough check, partly by presenting explicit formulas 
	(up to $o (\hbar^6, \ln (\Lambda))$ for arbitrary $N$) and
	partly by performing computer experiment (for $N=2,3,4$ where we check that
	the duality holds with higher precision but formulas are too lengthy to
	be manifestly written down).
	A general proof of the conjecture is still missing.

	\section {Outline}
	The conjecture itself and the way to check it are described in detail in \cite {MMBS1}. In this
	section we recall the construction very briefly.
	\par The statement that is made is that the Nekrasov function with one of the regularization parameters
	set to zero is equal to the quantized SW prepotential.
	\eq {
		\mF_{Nek} (\e_1, 0, \Lambda) \Big |_{\e_1 = \hbar} = \mF_{SW} (\hbar, \Lambda),
	}
	where the prepotential is defined by equations
	\eqna {
		a_i = \Pi^\hbar_{A_i} (\lambda) && \nonumber \\
		-\frac{1}{4} \frac{\p \mF_{SW}}{\p a_i} = \Pi^{\hbar}_{B_i} (\lambda)  \nonumber &&,
	}
	and $\Pi$'s are the Bohr-Sommerfeld periods of the quantum SW differential, and
	$\lambda$'s are the roots of the polynomial that enters the Fourier transform of
	the Baxter equation (see below).
	\par The proof includes evaluation of $\Pi^\hbar_{A_i}$ and $\Pi^\hbar_{B_i}$,
	which can be expressed through $\Pi^0_{A_i}$ and $\Pi^0_{B_i}$ - the periods for
	the classical SW differential. While $\Pi^0_{A_i}$ are easy to obtain by direct integration
	\cite {MasuSuzu}, direct calculation of $\Pi^0_{B_i}$ is not so easy, and one, for example,
	can use the well-known fact, that for $\hbar = 0$ the conjecture is valid
	\eq {
		\mF_{Nek} (0, 0, \Lambda) = \mF_{SW} (0, \Lambda)
	}
	From which one deduces
	\eq {
		\Pi^0_{B_i} = -\frac{1}{4}\frac{\p \mF_{SW}}{\p a_i} (0,0,\Lambda)
	}
	Then one constructs $\Pi^\hbar_{B_i}$ and looks if the result coincides with
	$-\frac{1}{4}\frac{\p \mF_{Nek}}{\p a_i} (\hbar,0,\Lambda)$. The subtlety here is, that
	$\Pi^\hbar_B$ and $-\frac{1}{4}\frac{\p \mF_{Nek}}{\p a_i} (\hbar,0,\Lambda)$ are
	expressed in terms of different variables ($\lambda$'s and $a$'s), so one needs to
	express $a$'s in terms of $\lambda$'s.
 	\par Mnemonically, one can write the conjecture in the following form
	\eq {
		\Pi^\hbar_B \lb \cmO \Pi^0_A \rb = \cmO \lb \Pi^0_B \lb \Pi^0_A \rb \rb,
	}
	which contains the way to check it.
	
	\bigskip
	
	\par The rest of the paper is organized as follows. In sections 3 and 4 the SW and Nekrasov
	sides of the conjecture are described in more detail. Attention is paid to the questions
	of automatization of calculations. Section 5 presents checks themselves.
	
	\section {Seiberg-Witten side}
	\subsection {Outlook}
	The classical Seiberg-Witten prepotential $\mathcal{F}_{SW}$ is determined by equations 
	\eq {
		a_i = \oint_{A_i} \lambda_{SW} = \Pi^0_{A_i}
	}
	\eq {
		\nonumber
		-\frac{1}{4}\frac{\p \mathcal{F}_{SW}}{\p a_i} = \oint_{B_i} \lambda_{SW} = \Pi^0_{B_i},
	}
	where $\lambda_{SW}$ is the Seiberg-Witten differential $\lambda_{SW} = p dx$ defined
	on the spectral curve given by equation 
	\eq {
		K(p) + \gamma \cos(x) = 0,
		\label{Baxter_classical}
	}
	where $K = \sum_j u_j p^j = u_N \prod_j (p-\lambda_j)$ is a polynomial. $A_i$ and $B_i$ form the symplectic basis
	of 1-cycles on this curve.
	\par The original idea of \cite{MMBS1} is to substitute ``classical'' SW-differential
	$p dx$ with ``quantum'' differential $P dx$, where $P$ solves quantum version of
	(\ref {Baxter_classical}) that is  the Fourier transform of the Baxter equation
	\eq {
		\lb K(\frac{\hbar}{i} \p) + \gamma \cos(x) \rb e^{\frac{i}{\hbar} \int^x P dx} = 0,
	}
	where $P = p + O(\hbar)$.
	\par Periods of this quantized differential then define $\mathcal{F}_{SW}(\hbar)$ in the similar
	way
	\eq {
		a_i = \Pi^\hbar_{A_i} (\lambda), \ \ -\frac{1}{4}\frac{\p \mathcal{F}_{SW} (\hbar)}{\p a_i} =
			\Pi^\hbar_{B_i} (\lambda),
	}
	where the implicit dependency $\lambda (a)$ needs to be resolved from the first set of
	equations (for A-cycles) and substituted into the second (for B-cycles).
	
	\par From the above it is clear that in order to define the prepotential
we need to know only periods of $\lambda_{SW}$, not $\lambda_{SW}$ itself,
which means that we can add arbitrary exact terms to $\lambda_{SW}$ to simplify
the calculations. Further, the idea of \cite{MMBS1} is to represent the quantized SW differential
as some differential operator $\cmO$ acting on the classical one $P dx = \cmO \lb p dx \rb$. Then for the periods
one also gets
	\eq { \nonumber
		\Pi^\hbar_{A_i} = \cmO \Pi^0_{A_i},\ \ \Pi^\hbar_{B_i} = \cmO \Pi^0_{B_i},
	}
	where $\cmO = 1 + O(\hbar)$.
	
	\par Thus, to find the quantum SW periods one needs to do the following:
	\begin{itemize}
	 \item solve the Baxter equation for $P$ perturbatively,
	 \item simplify $P$ by adding full derivatives,
	 \item rewrite the resulting $P$ in the form $\cmO \lb p \rb$.
	\end{itemize}
	The rest of this section is devoted to the details of these steps.
	
	\subsection {Solving Baxter equation}
	The first step is to evaluate $P$ perturbatively. When one considers conjugation of $\p^n$
	with $e^{\frac{i}{\hbar} \int^x P dx}$, one finds
	\eqna {
		e^{-\frac{i}{\hbar} \int^x P dx} \p^n e^{\frac{i}{\hbar} \int^x P dx} = 
		\nonumber & P^n + \frac{\hbar}{i} \frac{1}{2} n (n-1) \dot{P} & \\
		& + \lb \frac{\hbar}{i} \rb^2 \lb \frac{1}{6}n (n-1)(n-2) P^{n-3} \ddot{P} +
		\frac{1}{24} \cdot 3 n(n-1)(n-2)(n-3) P^{n-4} \dot{P}^2\rb + \dots, & 
	}
	where summation at level $k$ goes over partitions (Young diagrams) of weight $k$. Namely,
	the contribution of the partition $\ok = (k_1, \dots, k_m)$ equals ($(\wk_1 \dots \wk_n)$ denotes
	the conjugated partition)
	\eq {
		\lb \frac{\hbar}{i} \rb^{\alpha_m} \frac{1}{\alpha_m!} \cdot \frac{\beta}{\gamma} 
		n \dots (n-\alpha_m) P^{(k_1)} \dots P^{(k_m)} \equiv 
		\lb \frac{\hbar}{i} \rb^{\alpha_m} n \dots (n-\alpha_m) P^{(k_1)} \dots P^{(k_m)} C (\ok)
		\nonumber,
	}
	where the equivalence is the definition of the coefficients $C(\ok)$ and
	\eqna {
		\nonumber \alpha_j = \sum_{i=1}^j (k_i + 1) & \beta = \prod_{i=1}^m C_{\alpha_i}^{k_i+1}
		& \gamma = \prod_{i=1}^n (\wk_i-\wk_{i-1})!
	}
	
	\par Hence, the Baxter equation becomes
	\eq {
		\sum_{\ok} \lb \frac{\hbar}{i} \rb^{\alpha_m} K^{(\alpha_m)} (P) C (\ok) + \gamma \cos(x) = 0 ,
	}
	where $K^{(i)}$ stands for the $i$-th derivative of $K$.
	Having this, one can calculate $P = p + \hbar p_1 + \dots p_n$, up to any desired order.
	
	\subsection {Simplification of $P$}
	Since we are interested in $A$ and $B$ periods of $P dx$, not just in $P$ itself, we
	can add to it terms, which are exact, in order to simplify it.
	\par A typical term in $P$ looks like
	\eq {
		\frac{K^{(i_1)}(p) \dots K^{(i_j)}(p)}{\lb K^{'}(p) \rb^k} V^{(j_1)}
		 \dots V^{(j_l)} \label{typical_term}
	}
	The simplified form is such that the degree of $P$ in $\frac{1}{K^{'}}$ is the
	smallest. The motivation for this definition is that we want to rewrite our $P$ as some
	differential operator $\cmO$ acting on $p$, and the order of this operator is
	roughly speaking half of the degree of $P$ in $\frac{1}{K^{'}}$.
	\par We suggest the following ansatz for exact terms: they are themselves
	the derivatives of something of the form (\ref{typical_term}).
	\par
	To get the idea of how to simplify $P$ one should notice that among the terms of our
	ansatz the one which comes
	from differentiation of $1/K^{'}$ gives the biggest contribution to the degree of $P$. Indeed,
	compare these two lines
	\eq {
		\frac{\p}{\p x} \frac{1}{(K^{'})^k} = -\frac{K^{''} V^{'}}{(K^{'})^{k+2}}
	}
	\eq {
		\nonumber
		\frac{\p}{\p x} K^{(i)} = - K^{(i+1)} \frac{V^{'}}{K^{'}}
	}
	It is clear that the simplification procedure looks as follows: 
	\begin {itemize}
	 \item one looks for a term in $P$ that contains $\frac{K^{''}}{(K^{'})^n}V^{'}$ with
	 the biggest $n$,
	 \item subtracts from $P$ the corresponding full derivative
	 \item repeats the first two steps until there are terms in $P$ of that form.
	\end {itemize}
	
	\subsection {Perturbative answer $\rightarrow$ differential operator}
	Let's recall, that in order to find the periods of the quantum SW differential one
	needs to find the differential operator $\cmO$ such that $P = \cmO p$.
	It turns out, that up to $\hbar^4$ the simplified form of $P$ contains only even
	derivatives of $V$, which for $V = \cos(x)$ are proportional to $V$ itself.
	These are good news, since the differential operator $\cmO$ can be
	composed from the following elementary differential operators.
	\eq {
		\mathcal{D}_\gamma = \gamma \frac{\p}{\p \gamma}
	}
	\eq {
		\nonumber
		\mathcal{D}_{i} = \sum_{j \geq i} \frac{j!}{(j-i)!} u_j \frac{\p}{\p u_{j-i}} =
		i! \sum_{j \geq i} C^j_i u_j \frac{\p}{\p u_{j-i}}
	}
 The following identities are helpful
	\eqna {
		\mathcal{D}_\gamma (p) = -\frac{\gamma \cos(x)}{K^{'}} = -\frac{V}{K^{'}} &&
			\mathcal{D}_i (p) = -\frac{K^{(i)}}{K^{'}} \\
		\mathcal{D}_\gamma (K^{(j)}) = -\frac{K^{(j+1)} V}{K^{'}} &&
			\mathcal{D}_i (K^{(j)}) = K^{(j+i)} - \frac{K^{(j+1)} K^{(i)}}{K^{'}}
	}
	\par One can see, that the term of the highest degree in $\frac{1}{K^{'}}$ in the result
	of the action of differential operator $\D_{i_1} \cdots \D_{i_n} (\D_\gamma)^m$ 
	on $p$ is proportional to
	\eq {
		\D_{i_1} \cdots \D_{i_n} (\D_\gamma)^m (p)\Big{|}_{highest\ degree} \sim
		\frac{V^m (K^{''})^{m-1}}{(K^{'})^{2m-1}} \frac{(K^{''})^n
			K^{(i_1)} \cdots K^{(i_n)}}{(K^{'})^{2n}} \label {highest_order}
	}
	So, the procedure of finding the differential operator looks as follows.
	\begin {itemize}
	 \item one finds a term of the form (\ref{highest_order}) in $P$,
	 \item deduces the differential operator $\D$, that generates such term,
	 \item subtracts the result of the action of $\D$ on $p$ from $P$,
	 \item repeats until $P$ is equal to zero.
	\end {itemize}
	Summing up all $\D$'s found during this procedure, one obtains $\cmO$.

	\par After this procedure is applied, one ends with the following expression for
	the differential operator
	\eq { \boxed {
		\cmO = 1 + \frac{\hbar^2}{2^3 \cdot 3} \mathcal{D}_2 \mathcal{D}_\gamma
		+\frac{7 h^4}{2^7 \cdot 3^2 \cdot 5} \lb  
		\D_2 \D_2 \D_\gamma \D_\gamma - \frac{2}{7}\D_4 \D_\gamma \D_\gamma
		- \frac{2}{7}\D_2 \D_2 \D_\gamma \rb +  \cmO^{(6)} + o \lb \hbar^6 \rb,
	}
	}
	and the formula for $\cmO^{(6)}$ will be written below, since its derivation includes
	few additional tricks.

	\par The important thing to stress is, that the resulting operator is not just an arbitrary
	operator in $u_i$ and $\gamma$, how it could in principle have happened, but is composed of
	$\D_i$ and $\D_\gamma$, and hence lies in the really strict class of differential operators.
	One may hope that the same simplification occurs
	in the case of gauge theory with matter (the XXX chain).

	\paragraph {the 6th order in $\hbar$}
	At the sixth order in $\hbar$ a new subtlety appears: simplified $P$ contains
	$(V^{(3)})^2 \sim \sin^2(x)$,
	so at first sight the set $\lb \D_\gamma,\ \D_i \rb$ is not sufficient to express $\cmO^{(6)}$.
	However, let's
	examine the situation in more detail.
	\par Consider the more general form of the classical Baxter equation
	\eq {
		K(p) + \frac{\gamma}{2} e^{ix} + \frac{\beta}{2} e^{-ix} = 0,
		\label{Bax_gen}
	}
	so the previous form corresponds to $\beta = \gamma$.
	\par It is convenient to introduce the following operators
	\eqna {
		\D_{\gamma +} = \gamma \frac{\p}{\p \gamma} + \beta \frac{\p}{\p \beta} \\
		\D_{\gamma -} = \gamma \frac{\p}{\p \gamma} - \beta \frac{\p}{\p \beta} \\		
	}
	\par It is obvious that
	\eqna {
		\lb \D_{\gamma +} \rb^n (p) \Big{|}_{\beta = \gamma} =
		\frac{\cos^n(x)(K^{''})^{n-1}}{(K^{'})^{2n-1}} + \cdots && \\
		\lb \D_{\gamma -} \rb^n (p) \Big{|}_{\beta = \gamma} =
		(i)^n\frac{\sin^n(x)(K^{''})^{n-1}}{(K^{'})^{2n-1}} + \cdots, &&
	}
	\par and in order to express $\cmO^{(6)}$ we need only
	\eq {
		\lb \D_{\gamma -} \rb^2 (p) \Big{|}_{\beta = \gamma} =
		- \frac{\cos(x)}{K^{'}} - \frac{\sin^2(x) K^{''}}{(K^{'})^3}
	}
	\par The less obvious thing is that
	\eq {
		\lb \gamma \frac{\p}{\p \gamma} - \beta \frac{\p}{\p \beta} \rb \Pi^0 = 0
	}
	Indeed, $\oint p dx = -\oint x dp$, and if one performs the shift $x \rightarrow x + i \ln(\gamma)$,
	(\ref{Bax_gen}) becomes
	\eq {
		K(p) + \frac{1}{2} e^{ix} + \frac{\beta \gamma}{2} e^{-ix} = 0,
	}
	and calculating $x(p)$ perturbatively one always gets a function of $\gamma \beta$, 
	which is mapped to zero by $\D_{\gamma -}$. Further, since
	\eq {
		[ \D_{\gamma -},\ \D_{\gamma +}] = \D_{\gamma -},
	}
	only those terms in $\cmO$ which are free of $\D_{\gamma -}$ give non-zero contribution, when acting on $\Pi^0$.
	And this means, that after finding $\cmO$ in terms of $\D_{\gamma_+}$ and $\D_{\gamma_-}$,
	terms with $\D_{\gamma_-}$ can be dropped out and
	$\D_{\gamma +}$ can be substituted by $\gamma \frac{\p}{\p \gamma}$.

	\par After all this is performed, one gets the following answer for $\cmO^{(6)}$
	\eqna {
		\nonumber \cmO^{(6)} = \frac{\hbar^6}{2^7 \cdot 3^3 \cdot 5 \cdot 7}
		\Big{[} \lb \frac{31}{8} (\D_2)^3  - \frac{15}{4} \D_2 \D_4 + \frac{1}{3} (\D_3)^2
		+ \D_6 \rb (\D_\gamma)^3 + &&\\
		+ \lb - \frac{15}{4} (\D_2)^3  + 2 \D_2 \D_4 -(\D_3)^2
		- \D_6 \rb (\D_\gamma)^2 + && \\
		\nonumber + \lb (\D_2)^3  - \D_2 \D_4 + \frac{2}{3} (\D_3)^2
		\rb \D_\gamma \Big{]} &&
	}

	\paragraph {$D_i$ in terms of roots}
	\par Since $A$ and $B$ periods are conveniently written in terms of roots of $K$, not its coefficients,
	it is necessary to express $D_i$, and hence $\cmO$ in terms of roots.
	It can be verified by direct check, that the following expression for $D_i$ holds
	\eq {
		D_i = - \sum_m \frac{K^{(i)} (\lambda_m)}{K^{'}(\lambda_m)}
				\frac{\p}{\p \lambda_m},
	}
	which is rather simple.

	\subsection {Classical A-periods in terms of roots}
	In \cite {MasuSuzu} the general expression for the classical A-periods was obtained
	\eq {
		\Pi^0_{A_i} = \lambda_i + \sum_{n=1}^\infty \frac{\lb \frac{1}{2} \rb_n \Lambda^{2nN}}{n! (2n)!}
			\lb \frac{\p}{\p \lambda_i} \rb^{2n-1} \prod_{k \neq i} (\lambda_k - \lambda_i)^{-2n}
	}
	Up to $o(\Lambda^{4N})$ one gets
	\eq {
		\Pi^0_{A_i} = \lambda_i - \frac{\Lambda^{2N}}{2} \frac{1}{\prod \lambda^2} \sum \frac{1}{\lambda}
		- \Lambda^{4N} \frac{1}{\prod \lambda^4} \lb \lb \sum \frac{1}{\lambda}\rb^3
		    + \frac{3}{4} \lb \sum \frac{1}{\lambda} \rb \lb \sum \frac{1}{\lambda^2} \rb
		    + \frac{1}{8} \sum \frac{1}{\lambda^3}
		    \rb,
	}
	where $\prod \frac{1}{\lambda^n}$ is a shorthand for $\prod_{k \neq i} \frac{1}{\lambda_{ik}^n}$
	and $\sum \frac{1}{\lambda^n}$ is a shorthand for $\sum_{k \neq i} \frac{1}{\lambda_{ik}^n}$.

	\bigskip

	\section {Nekrasov side}
	The definition of the Nekrasov function can be found in numerous papers, for example
	\cite{sygr} and \cite{MMBS1}. Here we give the definition for $\e_2 = 0$. 
	The Nekrasov function $\mF_{Nek}$ is equal to the sum of perturbative and instantonic
	contributions.
	\eq {
		\mF_{Nek} = \mF_{pert} + \mF_{inst}
	}
	
	\par For $\mF_{pert}$ no nice expression is available, but there is one for $\frac{\p \mF_{pert}}{\p a_i}$,
	which is sufficient for our needs.
	
	\eq {
		- \frac{\p \mF_{pert}}{\p a_i} = \sum_{j \neq i} 4 a_{ij} \left [ 
		\lb \ln \frac{a_{ij}}{\Lambda} - 1 \rb +
		\sum_{m=1}^\infty \frac{B_{2m}}{2m (2m-1)} \lb \frac{\e_1}{a_{ij}} \rb^{2m}
		\right ]
	}
	For such definition of $\mF_{pert}$, $\mF_{inst}$ should be defined as follows (this
	deviates from \cite{sygr} slightly, but we believe it is the matter of convention)
	\eq {
		\mF_{inst} = - 2\e_1 \e_2 \ln Z_{inst},
	}
	and $Z_{inst}$ is the sum over n-tuples of partitions
	\eq {
		Z_{inst} = \sum_n \lb \wLambda^{2N n} \sum_{\ok, |k| = n} Z_{inst}^{\ok} \rb
	}
	\eq {
		Z_{inst}^{\ok} = \prod_{nl} \prod_{ij} \frac{a_{nl} + \e_1(i-1) + \e_2(-j)}
				{a_{nl} + \e_1 (i-1-\widetilde{k}_{li}) + \e_2(k_{nj} - j)},
	}
	where
	\eq {
		\wLambda^N = \frac{1}{2 i^N} \Lambda^N,
	}
	and $\Lambda$ is precisely the $\Lambda$ from the Seiberg-Witten side of the duality.
	
	\par In two-instanton approximation, $Z_{inst}$ equals (indices in square brackets
	label the types of n-tuples of Young diagrams).
	\eqna {
		Z_{inst} = 1 + \lb \frac{\lambda}{2} \rb^2 Z_{[1]} + 
		               \lb \frac{\lambda}{2} \rb^4 \lb Z_{[2]} + Z_{[1,1]} + Z_{[1],[1]}\rb && \\
		Z_{[1]} = - \frac{1}{\e_1 \e_2} \sum_{i=1}^N R_i (a_i) && \\
		Z_{[2]} = \frac{1}{2 \e_1^2 \e_2 (\e_1 - \e_2)} \sum_{i=1}^N R_i (a_i) R_i (a_i + \e_2) && \\ 
		Z_{[1,1]} = - \frac{1}{2 \e_1 \e_2^2 (\e_1 - \e_2)} \sum_{i=1}^N R_i (a_i) R_i (a_i + \e_1) && \\
		Z_{[1],[1]} = \frac{1}{2 \e_1^2 \e_2^2} \sum_{i \neq j} R_i (a_i) R_j (a_j)
						\frac{a_{ij}^2 (a_{ij}^2 - \e^2)}
						{(a_{ij}^2 - \e_1^2)(a_{ij}^2 - \e_2^2)} && \\
		R_i (x) = \frac{1}{\prod_{j \neq i} (x - a_j)(x-a_j + \e)}
	}
	
	\par From $Z_{inst}$, $\mF_{inst} (\e_1, 0)$ can be readily obtained (now in $R_i (x)$ $\e_2$
	is also put to zero). 
	\eqna {
		\mF_{inst} = \frac{1}{2} \lambda^2 \sum_i R_i(a_i) +
		\lb \frac{\lambda}{2} \rb^4 \Big{[}
			\frac{1}{\e_1} \sum_i \lb R_i^2(a_i) \sum_{j \neq i} 
				\lb \frac{1}{a_{ij}} + \frac{1}{a_{ij} + \e_1} \rb \rb && \\
			-\frac{1}{\e_1^2} \sum_i R_i^2(a_i)
			+\frac{1}{\e_1^2} \sum_i R_i (a_i) R_i (a_i + \e_1)
			+ \sum_{i \neq j} R_i(a_i) R_j(a_j) \frac{1}{a_{ij}^2 - \e_1^2}
		\Big{]}
	}
	
	\section {Check of the conjecture}
	\par As was pointed out already in \cite {MMBS1}, the calculation of both the SW and Nekrasov sides of
	the duality can be easily computerized, and hence the duality can be checked up to any desired order in $\hbar$ and $\Lambda$
	for any given $N$. However, if one tries to attack generic $N$, calculations should be performed
	by hands, and for high orders in $\Lambda$ presentation of the results is not an easy task (mainly because formulas
	are very lengthy and hence non-illustrative). So, for the sake of simplicity, here we present explicit calculations for
	arbitrary $N$ only up to $o \lb \hbar^6, \ln(\Lambda) \rb$.
	
	\par Explicit computer checks that we have performed for a few small $N$ allow
	us to state that the duality holds at least with following precision
	\begin {itemize}
		\item N = 2: up to $o(\hbar^6, \Lambda^{6N})$ at least,
		\item N = 3: up to $o(\hbar^6, \Lambda^{4N})$ at least,
		\item N = 4: up to $o(\hbar^6, \Lambda^{2N})$ at least.
	\end {itemize}
		
	\subsection {The zeroeth order in $\Lambda^N$}
	\par Here we concentrate on checking the conjecture up to $o(\hbar^6, \ln (\Lambda))$ for
	arbitrary $N$. Formulas in this case are quite simple and the ideas of the check are easy to
	illustrate.
	\par First of all, both classical and quantum A-periods on the SW side are equal to the
	corresponding roots of $K$.
	\eq {
		\Pi^\hbar_{A_i} = \Pi^0_{A_i} = a_i = \lambda_i,
	}
	since $\cmO$ acts nontrivially only on $\Lambda$-dependent terms, which in this case are
	missing.
	\par Hence, in what follows we write all formulas in terms of $a$'s instead of $\lambda$'s and
	hope this will not cause any confusion.
	
	\par Recall that the identity we want to check is
	$$
		\Pi^\hbar_{B_i} = \cmO \Pi^0_{B_i},
	$$
	or, in other words
	\eq {
		 \frac{\p \mF_{pert}}{\p a_i} (\hbar, \Lambda) =
		\cmO \frac{\p \mF_{pert}}{\p a_i} (0, \Lambda)
	}
	
	\par Up to $o (\hbar^6)$ $\Pi^\hbar_{B_i}$ is equal to
	\eq {
	\Pi^\hbar_{B_i} = -\frac{1}{4} \frac{\p \mF^{pert}}{\p a_i}(\hbar, \Lambda) =
		\sum_{j \neq i} \left [ a_{ij} \lb \ln \frac{a_{ij}}{\Lambda} - 1 \rb
			+ \frac{\hbar^2}{12} \frac{1}{a_{ij}^2}
			- \frac{\hbar^4}{360} \frac{1}{a_{ij}^4}
			+ \frac{\hbar^6}{1260} \frac{1}{a_{ij}^6}
		\right ]
	}
	$$
		\Pi^0_{B_i} = -\frac{1}{4} \frac{\p \mF^{pert}}{\p a_i}(0, \Lambda) =
		\sum_{j \neq i}  a_{ij} \lb \ln \frac{a_{ij}}{\Lambda} - 1 \rb
	$$
	
	\par Since $\cmO$ acts nontrivially only on $\Lambda$-dependent part of $\Pi^0_{B_i}$ and 
	$\D_\gamma \ln \gamma = 1$, the terms in $\cmO$ with higher than one power
	of $\D_\gamma$ do not
	contribute, and we are in fact checking the following identity
	
	\eqna {\label {Conj}
		&& \sum_{j \neq i} \frac{\hbar^2}{12} \frac{1}{a_{ij}}
			- \frac{\hbar^4}{360} \frac{1}{a_{ij}^3}
			+ \frac{\hbar^6}{1260} \frac{1}{a_{ij}^5} = \\
		\nonumber 
		&& \left [ \frac{\hbar^2}{2^3 \cdot 3} \D_2 +
			- \frac{\hbar^4}{2^6 \cdot 3^2 \cdot 5} \D_2 \D_2
			+ \frac{\hbar^6}{2^7 \cdot 3^3 \cdot 5 \cdot 7}
				\lb (\D_2)^3 - \D_2 \D_4 + \frac{2}{3} (\D_3)^2 \rb
		\right ] \lb -\frac{1}{N} \sum_{j \neq i} a_{ij} \rb,
	}
	since $\D_\gamma \lb  \ln \frac{1}{\Lambda} \rb = -\frac{1}{N}$.
	
	\par Naturally, the check splits into 3 checks for $\hbar^2$, $\hbar^4$ and $\hbar^6$,
	respectively. In what follows for each order in $\hbar$ first it is shown that the structure of both sides
	of the equality (\ref {Conj}) is the same and then that the coefficients do coincide.
	
	\paragraph {$\hbar^2$:}
	\eq {
		\D_2 \Pi^0 = -\sum_{j \neq i} \lb \sum_{k \neq i} \frac{2}{a_{ik}} -
			\sum_{k \neq j} \frac{2}{a_{jk}} \rb = -\sum_{k \neq i} 2\frac{(N-1) +1}{a_{ik}}
	}
	and 
	\eq {
		\frac{2}{2^3 \cdot 3} = \frac{1}{12}
	}
	\paragraph {$\hbar^4$:}
	\eq {
		\D_2 \D_2 \Pi^0 = -4N \sum_{j \neq i} \frac{1}{a_{ij}^2} \lb \sum_{k \neq i} \frac{1}{a_{ik}} -
			\sum_{k \neq j} \frac{1}{a_{jk}} \rb = -\sum_{j \neq i} \frac{8N}{a_{ij}^3},
	}
	since
	\eq {
		\sum_{k \neq i} \frac{1}{a_{ik}} - \sum_{k \neq j} \frac{1}{a_{jk}} = 
		\frac{2}{a_{ij}} + \sum_{k \neq i,j} \frac{a_{ji}}{a_{ik}a_{jk}},
	}
	and
	\eq {
		\sum_{j \neq i} \sum_{k \neq i,j} \frac{1}{a_{ij} a_{jk} a_{ki}} = 
		\sum_{j \neq i} \sum_{k \neq i,j} \frac{-1}{a_{ji} a_{kj} a_{ik}} =
		\sum_{k \neq i} \sum_{j \neq i,k} \frac{-1}{a_{ji} a_{kj} a_{ik}} = 0
	}
	The corresponding coefficient
	\eq {
		-\frac{8}{2^6 3^2 5} = -\frac{1}{360},
	}
	as it should be.
	
	\paragraph {$\hbar^6$:}
	Again, the stategy is to rewrite the r.h.s. in terms of 'nested sums',
	that is the sums in which all summation indices are mutually different, namely
	\eq {
		(\D_2)^3 \frac{1}{N}\sum_{j \neq i} a_{ij} = -96 \sum_{j \neq i} \frac{1}{a_{ij}^5} + 48 \sum_{j \neq i} \sum_{k \neq i,j} \frac{1}{a_{ij}^3 a_{ik}a_{jk}}
	}
	\eqna {
		\frac{2}{3} (\D_3)^2 \frac{1}{N}\sum_{j \neq i} a_{ij} = &
		-24 \sum_{j \neq i} \sum_{k \neq i,j} \frac{1}{a_{ij}^3 a_{ik}}
		\lb \frac{1}{a_{ik}} + \frac{1}{a_{jk}}\rb &  \nonumber \\
		& -24 \sum_{j \neq i} \sum_{k \neq i,j} \sum_{l \neq i,j,k} 
			\frac{1}{a_{ij}^2 a_{ik} a_{il}} \lb \frac{1}{a_{ij}} + \frac{1}{a_{ik}}\rb
			- \frac{1}{a_{ij}^2 a_{ik} a_{jl}} \lb \frac{1}{a_{ji}} + \frac{1}{a_{jk}}\rb & \\
		& - 12 \sum_{i \neq j} \sum_{k \neq i,j} \sum_{n \neq i,j,k} \sum_{l \neq i,j,k,n}
			\frac{1}{a_{ij}^2 a_{ik}} \lb \frac{1}{a_{in}a_{il}} -
								\frac{1}{a_{jn}a_{jl}}\rb & \nonumber
	}
	
	\eqna {
		\D_2 \D_4 \frac{1}{N}\sum_{j \neq i} a_{ij} = &
		- 16 \sum_{j \neq i} \sum_{k \neq i,j} \sum_{l \neq i,j,k} \sum_{p \neq i,j,k,l}
		\frac{1}{a_{ij}^2} \lb \frac{1}{a_{ik} a_{il} a_{ip}} - \frac{1}{a_{jk} a_{jl} a_{jp}}\rb & \\
		& -48 \sum_{j \neq i} \sum_{k \neq i,j} \sum_{p \neq i,j,k} \frac{1}{a_{ij}^3}
			\lb \frac{1}{a_{ik} a_{ip}} + \frac{1}{a_{jk} a_{jp}} \rb
	}
	\par First of all, let's note, that
	\eq {
		\frac{96}{2^7 \cdot 3^3 \cdot 5 \cdot 7} = \frac{1}{1260},
	}
	so the coefficient is correct.
	
	\par To complete the check, one must see, that double, triple and quadruple sums are zero.
	Here we do not write all of them, just a few examples to illustrate the idea, since
	the only trick one needs to use is relabeling of indices in sums. More precisely,
	one should represent given expression as a sum of 'cyclic sums', that is sums of the form
	$$
		\sum_{i_1 \neq i} \dots \sum_{i_{n} \neq i..i_{n-1}} \frac{1}{a_{i i_1} a_{i_1 i_2} \dots a_{i_n i}},
	$$
	and these cyclic sums can be shown to be equal to zero by noticing that such sum does not
	change under permutation  of indices $i_1 \dots i_n$ and then summing over all permutations.
	
	\bigskip
	
	\paragraph {Examples.}
	\par For the double sum we have
	\eq {
		\sum_{j \neq i} \sum_{k \neq i,j} \frac{1}{a_{ij}^3 a_{ik}}
			\lb \frac{1}{a_{ik}} - \frac{1}{a_{jk}}\rb =
		\sum_{j \neq i} \sum_{k \neq i,j} \frac{1}{a_{ij}^2 a_{ik}^2 a_{jk}} = 0,
	}
	due to the antisymmery of the summand under the exchange $j \leftrightarrow k$.
	\par For one of the quadruple sums we have (we omit the sum symbols to simplify formulas)
	\eq {
		\frac{1}{a_{ij}^2 a_{ik}} \lb \frac{1}{a_{in}a_{il}} -
		\frac{1}{a_{jn}a_{jl}}\rb \sim \frac{1}{a_{ij} a_{ik} a_{in} a_{jl} a_{jn}} +
		\frac{1}{a_{ij} a_{ik} a_{in} a_{jl} a_{il}},
	}
	where $'\sim'$ here and in what follows means that sometimes $\frac{1}{2}$ appears, which
	is inessential since we are showing that these sums are equal to zero.
	\par The second sum is zero because it contains the cycle $a_{ij} a_{jl} a_{li}$. In the first term
	we can exchange $l \leftrightarrow j$ and add this to  the result of exchange
	$l \leftrightarrow n$
	\eq {
		\frac{1}{a_{ij} a_{ik} a_{in} a_{jl} a_{jn}} \sim
		\frac{1}{a_{ik} a_{ij} a_{jn} a_{ni}} \lb \frac{1}{a_{jl}} - \frac{1}{a_{nl}} \rb \sim
		\frac{1}{a_{ik} a_{ij} a_{jl} a_{ln} a_{ni}} \sim
	}
	\eq {
		\nonumber
		\frac{1}{a_{ik} a_{ij} a_{jl} a_{ln} a_{ni}} +
		\frac{1}{a_{ik} a_{il} a_{lj} a_{jn} a_{ni}} +
		\frac{1}{a_{ik} a_{ij} a_{jn} a_{nl} a_{li}}  = 0		
	}

	\section {Conclusion}
	\par In this paper we made the next step in proving the conjecture that
	the quantized SW prepotential
	is equal to the Nekrasov function $\mF_{Nek}$ with vanishing $\e_2$. We considered the case
	of pure gauge $SU(N)$ theory. Explicit formulas for the check up to $o(\hbar^6, \ln \Lambda)$
	for arbitrary $N$ were
	presented, and some intermediate steps of the construction of the objects that appear at both
	sides of the duality, which are needed to computerize the check, were clarified. The quantization
	operator $\cmO$, which plays the central role at the SW side, was evaluated up to $o(\hbar^6)$.
	All these considerations allowed us to check the duality for non-zero instantonic
	numbers for a few small $N$ via computer.
	\par Still, our understanding of this duality is far from being clear. Many interesting
	questions remain, such as what is the structure of the coefficients of $\cmO$,
	what are the right terms to formulate the conjecture so that formulas
	become compact
	and whether the duality survives if one includes matter hypermultiplets.
	The work is in progress in these directions.
	
	\section*{Acknowledgements} Author is indebted to A.Mironov and A.Morozov for
	stimulating discussions. This work was partly supported by RFBR grant 07-02-00878,
	by joint grant 09-02-91005-ANF, by Russian Presidents Grant of Support for the 
	Scientific Schools NSh-3035.2008.2, by Federal Agency for Science and Innovations of
	Russian Federation under contract 02.740.11.5029 and by Dynasty Foundation.

\end {document}